# High thermoelectric performance of two-dimensional (PbTe)$_2$ layer

Caiyu Sheng, Dengdong Fan, Huijun Liu*

*Key Laboratory of Artificial Micro- and Nano-Structures of Ministry of Education and School of Physics and Technology, Wuhan University, Wuhan 430072, China*

The electronic, phonon and thermoelectric transport properties of (PbTe)$_2$ layer are systematically investigated by using first-principles pseudopotential method and Boltzmann transport equation. Our calculations demonstrate that there is a valley degeneracy of six for the top valence band, which leads to larger carrier concentration and thus higher electrical conductivity without obvious reduction in the Seebeck coefficient. Moreover, the intrinsic van der Waals interactions between neighboring Pb layers induce additional phonon scattering and thus ultrasmall lattice thermal conductivity. As a consequence, a maximum $p$-type $ZT$ value of 2.9 can be achieved at 1000 K. Moreover, we find almost identical $n$- and $p$-type $ZT$ in the temperature range from 300 K to 800 K.

Thermoelectric (TE) materials can directly convert heat into electricity and thus attract much attention in the science community due to the increasing environmental pollution and energy crisis [1]. The conversion efficiency of a TE material can be determined by the dimensionless figure-of-merit $ZT = S^2\sigma T/(\kappa_e + \kappa_l)$ [2], where $S$, $\sigma$, $T$, $\kappa_e$, $\kappa_l$ are the Seebeck coefficient, the electrical conductivity, the absolute temperature, the electronic thermal conductivity and the lattice thermal conductivity, respectively. A good TE material is expected to have a high $ZT$ value, which requires one to maximize the power factor ($PF = S^2\sigma$) and simultaneously minimize the thermal conductivity ($\kappa = \kappa_e + \kappa_l$) as much as possible. Unfortunately, it is usually very difficult to do so because almost all the transport coefficients in the $ZT$ expression are coupled with each other [3]. For decades, great efforts have been devoted to enhance the $ZT$ values, such as preparing nanocomposite materials [4], tuning electronic bands [5, 6], forming resonant levels [7], building superlattices [8], employing hierarchical architecture [9] and so on. Of course, good TE materials should have suitable band gaps and small thermal conductivities as a premise.

---

* Author to whom correspondence should be addressed. Electronic mail: phlhj@whu.edu.cn

Among many good TE materials, the PbTe alloy was found to exhibit very favorable thermoelectric performance due to intrinsically lower thermal conductivity and larger power factor caused by band convergence [5]. Such a pioneering study stimulates a lot of subsequent works to further enhance its $ZT$ values [10, 11, 12, 13]. Recently, Sa *et al.* proposed a series of two-dimensional (2D) group-IV chalcogenides (AX)$_2$ with a novel stacking order of X–A–A–X (X = Se, Te and A = Si, Ge, Sn, Pb) [14]. Taking the (PbTe)$_2$ layer as an example, the Pb and Te atoms are covalently bonded while weak van der Waals (vdW) interactions exist between neighboring Pb layers. Such mixed chemical bonding usually suggests very favorable thermoelectric performance, as previously found for the Cr$_2$Ge$_2$Te$_6$ compound [15]. It is thus natural to ask whether the 2D (PbTe)$_2$ could have higher $ZT$ values, as also governed by the well-known effect of low-dimensionalization [16, 17]. In this work, first-principles calculations and Boltzmann transport theory are combined to accurately predict the electronic, phonon and thermoelectric transport properties of (PbTe)$_2$ layer. We shall see that a maximum $ZT$ value of 2.9 can be achieved at 1000 K. Interestingly, we find almost identical *n*- and *p*-type $ZT$ in the temperature range from 300 K to 800 K, which is highly desired in the construction of thermoelectric devices.

The electronic properties of (PbTe)$_2$ layer are calculated within the framework of density functional theory (DFT) [18, 19] by using the projector augmented wave (PAW) method, which is coded in the so-called Vienna *ab-initio* simulation package (VASP) [20]. The exchange-correlation energy is in the form of Perdew-Burke-Ernzerhof (PBE) with the generalized gradient approximation (GGA) [21]. To accurately predict the band gap, we also consider the hybrid density functional in the form of Heyd-Scuseria-Ernzerhof (HSE) [22]. To derive the electronic transport coefficients, we adopt the semiclassical Boltzmann transport equation [23] and the deformation potential (DP) theory [24]. A vacuum thickness of 25 Å is used to minimize the interaction between the layer and its periodic images. The lattice thermal conductivity is determined by using the supercell approach and solving the Peierls-Boltzmann transport equation, as implemented in the so-called ShengBTE code [25]. The required second-order interatomic force constants (IFCs) are obtained using finite displacement method embedded in the PHONOPY package. The third-order IFCs are generated by considering interactions up to fourth nearest neighbors.

As illustrated in Figure 1, the 2D $(PbTe)_2$ crystallizes in a layered hexagonal structure with space group of $P\bar{3}m1$. The optimized lattice constant is 4.45 Å. The system forms stacks in the sequence of Te–Pb–Pb–Te, with covalently bonded Pb–Te and vdW connected Pb–Pb. Note that the Tkatchenko-Scheffler method is employed for the consideration of vdW interactions, which can cause additional phonon scattering as will be discussed later. Unlike graphene, we see from Fig. 1(a) that the $(PbTe)_2$ exhibits a bucking distance of $\Delta=1.64$ Å, and there is a vdW distance of $h=2.75$ Å between neighboring Pb layers. The calculated bond lengths are 3.03 Å and 3.75 Å for the Pb–Te and Pb–Pb bonds, respectively. All these values are in good agreement with previous theoretical result [14] and further confirms the reliability of our approach. On the other hand, we see obvious threefold and sixfold rotational symmetries along the *z*-direction for the Pb and Te sites, respectively (Fig. 1(b)). It is expected that such a configuration characteristic may lead unique band shape and therefore influence the electronic transport properties [26].

Figure 2(a) plots the energy band structure of $(PbTe)_2$ layer. By using the conventional PBE functional, we see that the system exhibit an indirect band gap of 0.38 eV. The conduction band minimum (CBM) is located at the Γ point, while the valence band maximum (VBM) appears between the Γ and K points. As standard DFT tends to underestimate the band gap seriously, we adopt the hybrid functional in the form of HSE which can overcome such deficiency very effectively. By including the effect of spin-orbit coupling (SOC), the HSE-calculated band gap is enlarged to 0.77 eV, although with very similar dispersions around the Fermi level. It should be mentioned that along the ΓM direction, there is a valence band extremum (VBE) with almost the same energy as that of VBM which is reminiscent of band degeneracy. As a consequence, we see from the isoenergy lines of the top valence bands (Fig. 2(b)) that there is a degeneracy of six originated from the threefold rotational symmetry mentioned above. Such kind of multi-valley structure is beneficial for increasing the electrical conductivity without obvious reduction in the Seebeck coefficient due to induced larger carrier concentration and bigger density of state (DOS) effective mass [27]. It is therefore expected that the *p*-type (PbTe)$_2$ layer may exhibit higher power factor than the *n*-type system, which is indeed the case as plotted in Fig. 2(c) as a function of carrier concentration. For example, the room temperature power factor along the *x*-direction is as high as $3 \times 10^{-3}$ W/mK$^2$ for the *p*-type system, which is comparable with that of good thermoelectric materials such as Bi$_2$Te$_3$ [28]. Besides, we find obvious anisotropy of *p*-type power factor ($PF_x > PF_y$), which is not the case for the *n*-type system where identical $PF$ is found along the *x*- and *y*-directions. This is reasonable since the multi-valley structures shown in Fig. 2(b) indicate anisotropic top valence bands, while the CBM is only located at the Γ point. Note that the relaxation time ($\tau$) should be included to calculate the power factor. Within the framework of Boltzmann theory, it is well known that the evaluation of the electrical conductivity ($\sigma$) and the electronic thermal conductivity ($\kappa_e$) depends on the value of $\tau$, while the Seebeck coefficient ($S$) does not. Instead of using a constant relaxation time, here we adopt the DP theory assuming the single parabolic band (SPB) model [29], which is given by $\tau = \dfrac{2\hbar^3 C}{3 k_B T m^*_{dos} E_1^2}$ for 2D systems. Here $C$ is the elastic module, $T$ is the temperature, $m^*_{dos}$ is the DOS effective mass, and $E_1$ is the DP constant representing the shift of VBM and CBM per unit strain. The calculated room temperature relaxation time and the related parameters are summarized in Table 1. We see that $m^*_{dos}$ of *n*-type (PbTe)$_2$ layer is only 0.093 $m_e$, significantly smaller than that of *p*-type system of 0.62 $m_e$. As a consequence, we expect obviously larger relaxation time and thus higher electrical conductivity for electrons. This is however not the case for the Seebeck coefficient where we find obviously larger absolute value for holes. Such a different behavior leads to $PF_p > PF_n$ along the *x*-direction while $PF_n > PF_p$ along the *y*-direction, as indicated in Fig. 2(c).

Figure 3(a) plots the phonon dispersion relations of the (PbTe)$_2$ layer as obtained from the second-order IFCs, where no imaginary frequency is found suggesting the dynamical stability of the system. As the primitive cell contains four atoms, we see 12 phonon branches among which three optical ones mixed with the acoustic ones in the frequency range from 20 to 39 cm$^{-1}$. Such kind of hybridization can be found in many systems with intrinsically low thermal conductivities [30]. Moreover, there is a large phonon gap between 50 and 100 cm$^{-1}$, and the cutoff frequency of 140 cm$^{-1}$ is comparable to those of reported good TE materials such as Bi$_2$Te$_3$ (150 cm$^{-1}$) [31, 32]. At frequency higher than 100 cm$^{-1}$, we see six flat optical branches which are usually believed to be heat insulating due to very small group velocity and phonon relaxation time [33]. All these observations suggest that the (PbTe)$_2$ layer should have rather low lattice thermal conductivity, as previously found for the BiOCuSe compound [34]. Fig. 3(b) plots the phonon DOS (PDOS) of (PbTe)$_2$ layer as a function of frequency, where we see that the phonon modes of Pb atoms play a major role in the low-frequency region, while those of Te atoms dominate the high-frequency region. Such kind of mismatch in the PDOS suggests weaker vibrational interactions between the Pb and Te atoms, and the phonons can be easily scattered which induces inefficient heat transfer [35, 36]. On the other hand, the vdW interactions between neighboring Pb layers give additional interface phonon scattering and further reduce the lattice thermal conductivity [37]. Indeed, we see from Fig. 3(c) that the $\kappa_l$ of (PbTe)$_2$ layer in the vicinity of 1000 K is as small as 0.21 and 0.20 W/mK along the *x*- and *y*-directions, respectively. Note that the thermal conductivity of low-dimensional system is somehow arbitrary and here the value is given with respect to a vacuum distance of 25 Å. Considering the bucking distance ($\Delta$) and the vdW distance ($h$) mentioned above, we adopt a realistic thickness of 4.39 Å ($\Delta+h$) and the lattice thermal conductivity of (PbTe)$_2$ layer is recalculated to be 1.20 and 1.14 W/mK for the *x*- and *y*-directions, respectively. These renormalized values are rather small which suggest very favorable TE performance. To distinguish the contribution of phonons with different frequencies, we plot in Fig. 3(d) the cumulative lattice thermal conductivity ($\kappa_{cumu}$) as a function of cutoff frequency [38]. As expected, the low-frequency phonons dominate the thermal transport and we see that ~90% of the total lattice thermal conductivity are contributed by phonons with frequency below 50 cm$^{-1}$, which is consistent with their relatively larger group velocity. In particular, we find that phonons in the frequency region from 20 to 39 cm$^{-1}$ give almost half contribution to the lattice thermal conductivity, which is reasonable since the acoustic and optical phonon modes mixed with each other as mentioned above.

With all the electronic and phonon transport coefficients available, we can now evaluate the *ZT* values of (PbTe)$_2$ layer. As shown in Figure 4, we see obvious anisotropy of the *p*-type *ZT* which is consistent with different power factors along the *x*- and *y*-directions (see Fig. 3(c)) and is rooted in the anisotropy of top valence bands. At temperature of 1000 K, a maximum *p*-type *ZT* of 2.9 (2.6) can be reached along the *x*-direction (*y*-direction) with optimized carrier concentration of 5.4×10$^{19}$ cm$^{-3}$ (5.8×10$^{19}$ cm$^{-3}$). Such record high values suggest that the energy conversion efficiency of (PbTe)$_2$ layer can be comparable to the traditional power generation methods. In principle, such favorable TE performance is originated from the intrinsically low thermal conductivity of (PbTe)$_2$ layer, as well as enhanced electrical conductivity due to the multi-valley band structure. On the other hand, it is interesting to find that the *ZT* values of *p*- and *n*-type systems along the *x*-direction are very close to each other in the temperature range from 300 K to 800 K, which is very desirable for thermoelectric devices where TE modules with comparable *p*- and *n*-type efficiencies are technically wanted.

**Acknowledgments**

We thank financial support from the National Natural Science Foundation of China

(Grant Nos. 51772220 and 11574236).

**Table 1** The calculated room temperature relaxation time and the related parameters of (PbTe)$_2$ layer.

| direction | carrier type | $C$ (eV/Å$^2$) | $m^*_{dos}$ (m$_0$) | $E_1$ (eV) | $\tau$ (s) |
|---|---|---|---|---|---|
| $x$ | electron | 3.29 | 0.093 | −7.13 | 9.00×10$^{-14}$ |
|  | hole | 3.29 | 0.62 | −5.23 | 2.52×10$^{-14}$ |
| $y$ | electron | 3.28 | 0.093 | −7.13 | 8.97×10$^{-14}$ |
|  | hole | 3.28 | 0.62 | −5.97 | 1.93×10$^{-14}$ |

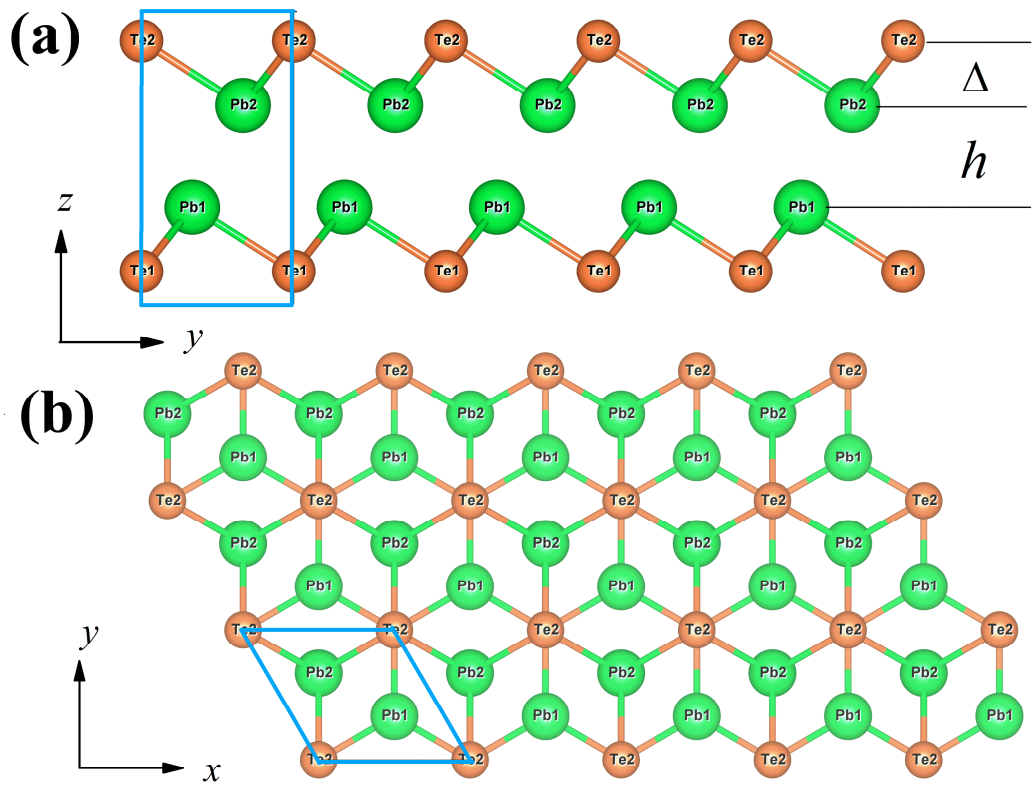

**Figure 1** Ball-and-stick model of (PbTe)₂ layer (a) side-view, and (b) top-view. The bucking and vdw distances are labeled as $\Delta$ and $h$, respectively.

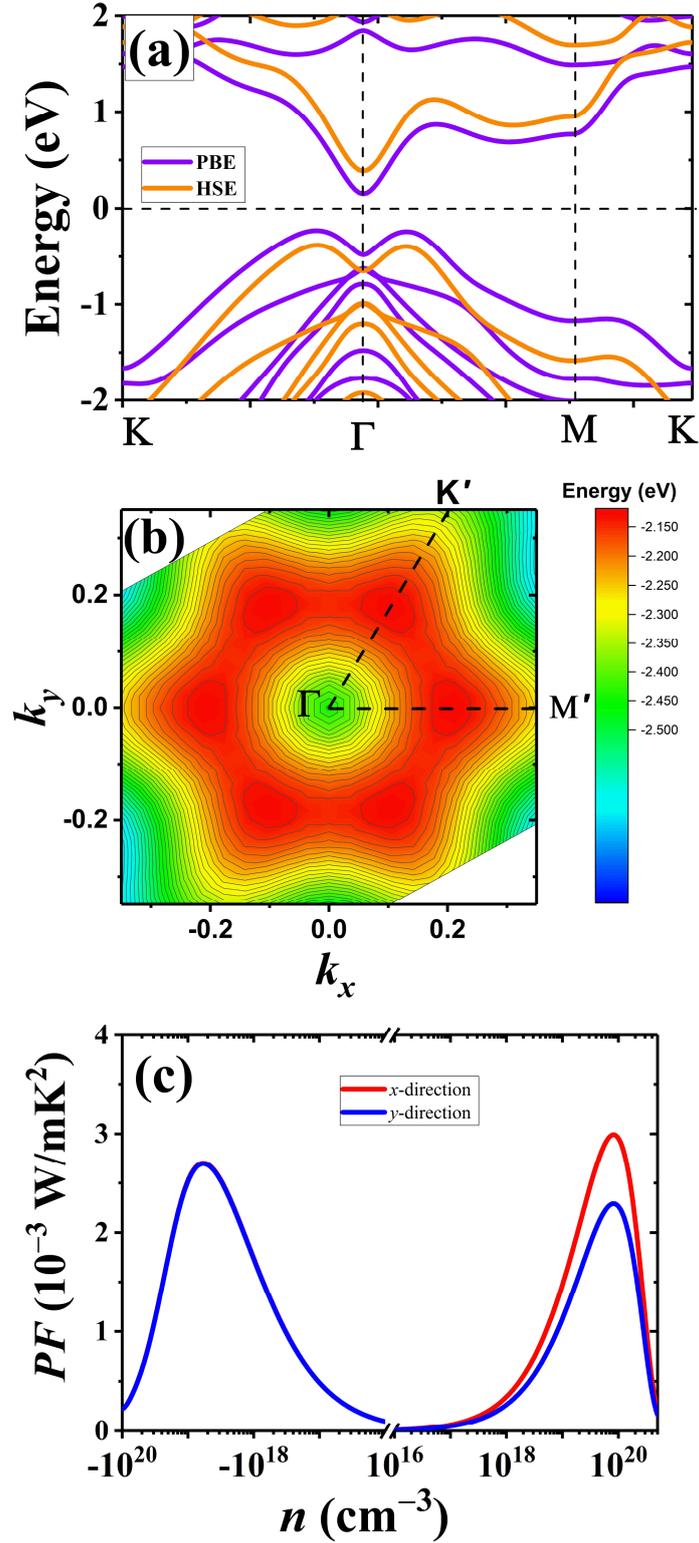

**Figure 2** (a) The electronic band structures of (PbTe)$_2$ layer. (b) The isoenergy lines of top valence bands. (c) The room temperature power factor as a function of carrier concentration.

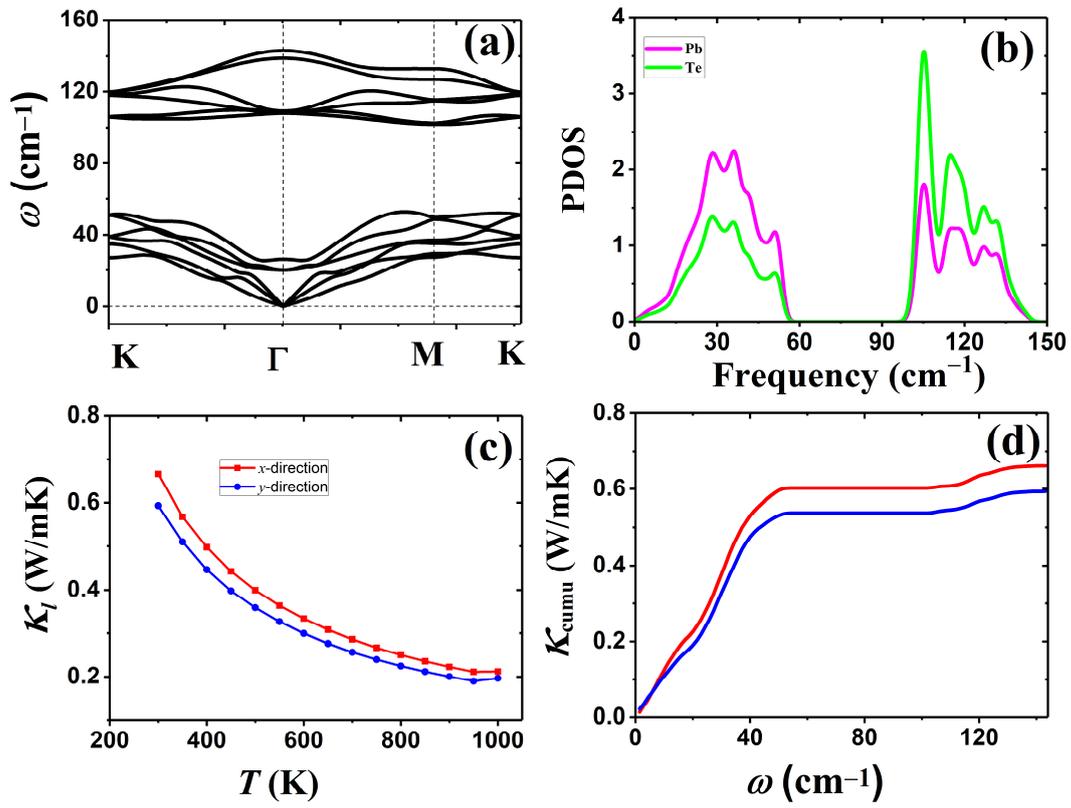

**Figure 3** (a) The phonon dispersion relations of (PbTe)$_2$ layer. (b) The corresponding PDOS. (c) The lattice thermal conductivity as a function of temperature. (d) The cumulative lattice thermal conductivity at 300 K, plotted as a function of cutoff phonon frequency.

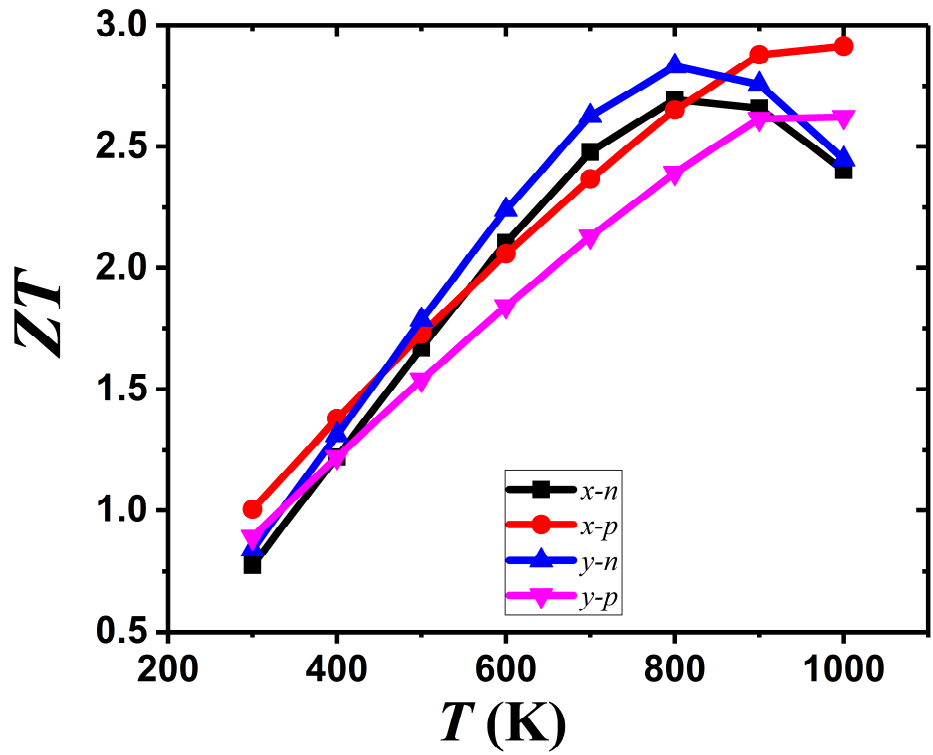

**Figure 4** Temperature dependent $ZT$ values of $(PbTe)_2$ layer. For comparison, the results of *n*- and *p*-type systems are both shown along the *x*- and *y*- directions.

**Reference**

[1] C. B. Vining, Nat. Mater. **8**, 83 (2009).

[2] H. J. Goldsmid, Introduction to Thermoelectricity (Springer, New York, 2009).

[3] G. Slack, in CRC Handbook of Thermoelectrics (Ed: D. M. Rowe), (CRC Press, Boca Raton, FL, 1995).

[4] M. S. Dresselhuus, G. Chen, M. Y. Tang, R. G. Yang, H. Lee, D. Z. Wang, Z. F. Ren, J. P. Fleurial and P. Gogna, Adv. Mater. **19**, 1043 (2007).

[5] Y. Z. Pei, X. Y. Shi, A. LaLonde, H. Wang, L. D. Chen and G. J. Snyder, Nature **473**, 66 (2011).

[6] W. Liu, X. J. Tan, K. Yin, H. J. Liu, X. F. Tang, J. Shi, Q. J. Zhang and C. Uher, Phys. Rev. Lett. **108**, 166601 (2012).

[ 7 ] J. P. Heremans, V. Jovovic, E. S. Toverer, A. Saramat, K. Kurosaki, A. Charoenphakdee, S. Yamanaka and G. J. Snyder, Science **321**, 554 (2008).

[8] G. Chen, Phys. Rev. B **57**, 14958 (1998).

[9] K. Biswas, J. Q. He, I. D. Blum, C.-I Wu, T. P. Hogan, D. N. Seidman, V. P. Dravid and M. G. Kanatzidis, Nature **489**, 414 (2012).

[10] H. J. Wu, L. D. Zhao, F. S. Zheng, D. Wu, Y. L. Pei, X. Tong, M.G. Kanatzidis and J. Q. He, Nat. Commun. **5**, 4515 (2014).

[11] Z. W. Chen, Z. Z. Jian, W. Li, Y. J. Chang, B. H. Ge, R. Hanus, J. Yang, Y. Chen, M. X. Huang, G. J. Snyder and Y. Z. Pei, Adv. Mater. **29**, 1606768 (2017).

[12] J. Cao, J. D. Querales-Flores, A. R. Murphy, S. Fahy and I. Savic, Phys. Rev. B **98**, 205202 (2018).

[13] H. T. Liu, Z. Y. Chen, C. Yin, B. Q. Zhou, B. Liu and R. Ang, Appl. Phys. A **125**, 225 (2019).

[14] B. S. Sa, Z. M. Sun and B. Wu, Nanoscale **8**, 1169 (2016).

[15] X. D. Tang, D. D. Fan, K. L. Peng, D. F. Yang, L. J. Guo, X. Lu, J. Y. Dai, G. Y. Wang, H. J. Liu and X. Y. Zhou, Chem. Mater. **29**, 7401 (2017).

[16] L. D. Hicks and M. S. Dresselhaus, Phys. Rev. B **47**, 12727 (1993).

[17] L. D. Hicks and M. S. Dresselhaus, Phys. Rev. B **47**, 16631 (1993).

[18] G. Kresse and J. Hafner, Phys. Rev. B **47**, 558 (1993).

[19] G. Kresse, and J. Hafner, Phys. Rev. B **49**, 14251 (1994).

[20] G. Kresse and J. Furthmüller, Comput. Mater. Sci. **6**, 15 (1996).

[21] J. P. Perdew, K. Burke and M. Ernzerhof, Phys. Rev. Lett. **77**, 3865 (1996).

[22] J. Heyd, G. E. Scuseria and M. Ernzerhof, J. Chem. Phys. **118**, 8207 (2003).

[23] G. K. H. Madsen and D. J. Singh, Comput. Phys. Commun. **175**, 67 (2006).

[24] J. Bardeen and W. Shockley, Phys. Rev. **80**, 72 (1950).

[25] W. Li, J. Carrete, N. A. Katcho and N. Mingo, Comput. Phys. Commun. **185**, 1747 (2014).

[26] Y. Li, J. Liu, Y-F Chen, F-N Wang, X-M Zhang, X-J Wang, J-C Li and C-L Wang, J. Materiomics **5**, 51 (2019).

[27] R. Q. Guo, X. J. Wang, Y. D. Kuang and B. L. Huang, Phys. Rev. B **92**, 115202 (2015).

[28] L. Cheng, H. J. Liu, J. Zhang, J. Wei, J. H. Liang, J. Shi and X. F. Tang, Phys. Rev. B **90**, 085118 (2014).

[29] D. M. Rowe, C. M. Bhandari, Modern thermoelectric, Reston Publishing Company, Inc.: Reston Virginia, pp. 26 (1983).

[30] D. Wee, B. Kozinsky, N. Marzari, and M. Fornari, Phys. Rev. B **81**, 045204 (2010).

[31] B. Qiu and X. L. Ruan, Phys. Rev. B **80**, 165203 (2009).

[32] B. Poudel, Q. Hao, Y. Ma, Y. C. Lan, A. Minnich, B. Yu, X. Yan, D. Z. Wang, A. Muto, D. Vashaee, X. Y. Chen, J. M. Liu, M. S. Dresselhaus, G. Chen and Z. F. Ren, Science **320**, 634 (2008).

[33] D. D. Fan, H. J. Liu, L. Cheng, J. Zhang, P. H. Jiang, J. Wei, J. H. Liang and J. Shi, Phys. Chem. Chem. Phys. **19**, 12913 (2017).

[34] S. K. Saha, Phys. Rev. B **92**, 041202 (2015).

[35] M. Hu, Y. H. Jing and X. L. Zhang, Phys. Rev. B **91**, 155408 (2015).

[36] G. Q. Ding, J. Carrete, W. Li, G. Y. Gao and K. L. Yao, Appl. Phys. Lett. **108**, 233902 (2016).

[37] F. Rieger, K. Kaiser, G. Bendt, V. Roddatis, P. Thiessen, S. Schulz and C. Jooss, J. Appl. Phys. **123**, 175108 (2018).

[38] Z. Rashid, A. S. Nissimagoudar and W. Li, Phys. Chem. Chem. Phys. **21**, 5679 (2019).